\documentclass[a4paper,12pt]{article}

\usepackage{amsmath,amsthm}
\usepackage{amsfonts}
\usepackage{xcolor}

\title{\textbf{Some comments on Gao beam model}}
\author{
  Jitka \textsc{Machalov\'{a}},\, Horym\'{i}r \textsc{Netuka} \\
  \normalsize {Department of Mathematical Analysis and Applications of Mathematics} \vspace*{-2pt} \\
  \normalsize {Faculty of Science, Palack\'{y} University, Olomouc, Czech Republic} \\
  \footnotesize \textit {jitka.machalova@upol.cz,\, horymir.netuka@upol.cz} \\
}
\date{\small \today}

\begin{document}

\maketitle

\begin{abstract}
In this small comment mathematical formulations concerning the nonlinear beam model published in \cite{gao96} are analyzed. The beam is subjected to vertical and axial loading (at its right end).
This nonlinear model can be used to study post-buckling problems.
Unfortunately, some inconsistency between pure bending and pure buckling problems was discovered by the authors of this comment. This is concerned with definition of an integral constant, which is not in \cite{gao96} strictly determined.
In this comment there is proposed the adjustment of of this constant which should solve the mentioned troubles. 
\end{abstract}
\bigskip

The Gao beam equation was firstly introduced in \cite{gao96}. This model uses the following assumptions
\begin{itemize}
  \item the Euler--Bernoulli hypothesis holds, i.e. plane cross sections perpendicular to the beam axis before bending remain plane and perpendicular after deformation, shear deformations are ignored,
  \item the material of the beam is isotropic, i.e. the Young's modulus $E$ is a constant,
  \item the beam has a uniform cross-section of a rectangular shape.
\end{itemize}

The beam is subjected to a vertical distributed load $q(x)$ and a horizontal
constant axial load $P$  at the end\, $x = L$\, which is positive in the negative $x$-direction (and vice versa).
The functional of potential energy of this beam was written in \cite{gao96} according to plane stress theory as

\begin{equation}\label{BPEnergy}
  \Pi (u,w)\, =\, \frac{1}{2} \int_{\Omega} (\sigma_x \epsilon_x + \sigma_y
                \epsilon_y)\hspace*{1pt} \mathrm{d}x \mathrm{d}y\hspace*{1pt} -
                \int^{L}_{0} {q} w\hspace*{1pt} \mathrm{d}x\hspace*{1pt} +
                \hspace*{1pt} P\hspace*{1pt} u(L),
\end{equation}
where
\begin{equation}\label{sigma}
  \left(
   \begin{array}{l}
    \sigma_x \\ \sigma_y
   \end{array}
  \right)\, =\,
  \frac{E}{1-\nu^2}\hspace*{1pt}
  \left(
   \begin{array}{cc}
    1 & \nu \\ \nu & 1
   \end{array}
  \right)
  \left(
   \begin{array}{l}
    \epsilon_x \\ \epsilon_y
   \end{array}
  \right),
\end{equation}
\begin{equation}
  \epsilon_{x} = u{'} -\, y w{''} +\, \frac{1}{2}\hspace*{1pt} (w{'})^2,\quad
  \epsilon_{y} = \frac{1}{2}\hspace*{1pt} (w{'})^2
\end{equation}
and\, $\Omega = [0, L] \times [-h, h]$.
Here $w(x)$ denotes the transverse displacement and $u(x)$ the horizontal displacement of the middle axis\, $y = 0$. As the cross-section of the beam is assumed to be constant and rectangular, area moment of inertia $I$ has the form
\begin{equation}\label{msp}
  I\, =\, \frac{2}{3}\, b\hspace*{1pt} h^3,
\end{equation}
where $b$ is width of the beam and $h$ is its half-thickness, then by using variational
methods we can obtain system of two nonlinear equations for stationary point of functional $\Pi(u,w)$:
\begin{eqnarray}
  \label{gaosystem1}
  u{''} +\hspace*{1pt} (1 + \nu)\hspace*{1pt} w{'} w{''}\!\! &=&\!\! 0, \\
  \label{gaosystem2}
  E I\hspace*{1pt} w{''''} -\hspace*{1pt} 2 h b E \left[ (1 + \nu)\
  \hspace*{1pt} (2 (w{'})^{2} + u{'})\hspace*{1pt} w{''} + \nu\hspace*{1pt} w{'} u{''} \right]\!\! &=&\!\! f,
\end{eqnarray}
where\, $f(x) = (1 - \nu^2)\hspace*{1pt} q(x)$ and\, $\nu > 0$\, denotes the Poisson ratio.
There is a little difference in expression of $I$ compared to paper \cite{gao96},
where the author used the expression\, $I = \frac{2}{3}\, h^3$\, as the width $b$ in \cite{gao96} was considered to be unit. In this paper the symbol $b$ is used due to physical units consistency and appears therefore also in (\ref{gaosystem2}).

Integrating (\ref{gaosystem1}), we have
\begin{equation}\label{du/dx}
  u{'}\, =\, -\hspace*{1pt} \frac{1}{2}\hspace*{2pt} (1 + \nu) (w{'})^{2}\hspace*{1pt} + C,
  %\hspace*{2pt} \frac{(1 - \nu^2)\hspace*{1pt} P}{2 \mathrm{h}\mathrm{b} E}
\end{equation}
where $C$ is an integral constant. Substituting (\ref{du/dx}) together with (\ref{gaosystem1}) in (\ref{gaosystem2}) we obtain
\begin{equation}\label{Gaobeam}
  E I\hspace*{1pt} w{''''}\hspace*{1pt} -\, 3\hspace*{1pt} h b\hspace*{1pt} E\hspace*{1pt} (1 - \nu^2)\hspace*{1pt} (w{'})^2 w{''}\hspace*{1pt} -\, 2\hspace*{1pt} h b\hspace*{1pt} E\hspace*{1pt} (1 + \nu)\hspace*{1pt} C\hspace*{1pt} w{''}\hspace*{1pt} =\, f.
\end{equation}

The approach in \cite{gao96} was based on setting of the integral constant $C$ in the form
\begin{equation}\label{C}
  C\, =\, -\hspace*{1pt} \frac{\lambda}{2 h b\hspace*{1pt} (1 + \nu)} \hspace*{1pt} .
\end{equation}
Consequently, in \cite{santos-gao12} the constant $\lambda$ was more specified as
\begin{equation}\label{lambda}
  \lambda\, =\, (1 + \nu) (1 - \nu^2)\, \frac{P}{E}\hspace*{1pt} .
\end{equation}
Hence the final beam equation reads as follows
\begin{equation}\label{GBfinal}
  E I\hspace*{1pt} w{''''}\hspace*{1pt} -\, E \alpha\hspace*{1pt} (w{'})^2 w{''}\hspace*{1pt} +\, (1 + \nu) (1 - \nu^2)\, P\hspace*{1pt} w{''}\hspace*{1pt} =\, f,
\end{equation}
where\, $\alpha = 3\hspace*{1pt} h b\hspace*{1pt} (1 - \nu^2)$\, and\,
$f = (1 - \nu^2)\hspace*{1pt} q$.

\bigskip
Now we can compare the Gao beam model with the classical Euler-Bernoulli beam (abbrev. as EB beam):
\begin{equation}\label{EBbeam}
  E I\hspace*{1pt} w{''''}\hspace*{1pt} +\, P\hspace*{1pt} w{''}\hspace*{1pt} =\, q.
\end{equation}

Hereafter let $E$ and $I$ be the same for both beams.

We can {take into account} two situations.
Firstly we consider the {\it pure bending problem}, i.e.\, $P = 0$, and let $q$ be the same for both beams. Then the EB beam equation reads as
\begin{equation}\label{EBpurebending}
  E I\hspace*{1pt} w{''''}\hspace*{1pt} =\, q
\end{equation}
and for Gao beam we have
\begin{equation}\label{GBpurebending}
  E I\hspace*{1pt} w{''''}\hspace*{1pt} -\, E \alpha\hspace*{1pt} (w{'})^2 w{''}\hspace*{1pt}
  =\, (1 - \nu^2)\hspace*{1pt} q.
\end{equation}
It is obvious that the final vertical load for Gao beam is smaller than for EB beam, because\,  $(1 - \nu^2) < 1$.
Dividing the last equation by\, $(1 - \nu^2)$\, we get the both beams under the same applied vertical load
\begin{equation}
  \widetilde{E} I\hspace*{1pt} w{''''}\hspace*{1pt} -\, \widetilde{E} \alpha\hspace*{1pt} (w{'})^2 w{''}\hspace*{1pt} =\, q,
\end{equation}
where\, $\widetilde{E} := E / (1 - \nu^2)$. It is clear that the Gao beam is tougher than EB beam because\, $\widetilde{E} > E$.

This fact was confirmed also by means of computational experiments (see e.g. \cite{machalova17}, \cite{machalova18}). The following simple example can be {analysed} without computations. Let the beam with simple support at both ends {be} given and let\, $q(x) = \mathrm{const} < 0$\, for all\, $x \in [0, L]$. The equation (\ref{GBpurebending}) can be rearranged using definition of $\alpha$ this way
\begin{equation}\label{GBpbre}
  E I\hspace*{1pt} w{''''}\hspace*{1pt}
  =\, (1 - \nu^2) \left[ q +\, 3\hspace*{1pt} h b\hspace*{1pt} E\hspace*{1pt} (w{'})^2 w{''} \right].
\end{equation}
The given $q$ implies evidently convex shape of $w(x)$, hence\, $w{''}(x) > 0$\, for all $x$ and therefore\, $\tilde{q}(x) := q +\, 3\hspace*{1pt} h b\hspace*{1pt} E\hspace*{1pt} ((w{'})^2 w{''})(x)$\, causes less bending (in the negative $y$-direction) than $q$. And moreover, the loading\, $(1 - \nu^2)\hspace*{1pt} \tilde{q}(x)$\, is once again smaller than the original loading $q$. All this implies that bending which can be obtained from (\ref{EBpurebending}) must be greater (in absolute values) than bending defined by (\ref{GBpurebending}).

\smallskip
Secondly, let us consider {\it pure buckling problem}, i.e.\, $q = 0$, and let $P$ be the same for both beams. Then we have for EB beam
\begin{equation}\label{EBpurebuckling}
  E I\hspace*{1pt} w{''''}\hspace*{1pt} + P \hspace*{1pt} w{''}\hspace*{1pt} =\, 0
\end{equation}
and for Gao beam
\begin{equation}\label{GBpurebuckling}
  E I\hspace*{1pt} w{''''}\hspace*{1pt} -\, E \alpha\hspace*{1pt} (w{'})^2 w{''}\hspace*{1pt} +
  \, (1 + \nu) (1 - \nu^2)\, P\hspace*{1pt} w{''}\hspace*{1pt} =\, 0.
\end{equation}
Now the idea is similar as in the pure bending problem. To have both beam under the equal axial load we can divide the last equation by\, $(1 + \nu) (1 - \nu^2)$\, and get
\begin{equation}
  \bar{E} I\hspace*{1pt} w{''''}\hspace*{1pt} -\, \bar{E} \alpha\hspace*{1pt} (w{'})^2 w{''}\hspace*{1pt} +\, P \hspace*{1pt} w{''}\hspace*{1pt} =\, 0,
\end{equation}
where
\begin{equation}
  \bar{E} := \frac{E}{(1 + \nu) (1 - \nu^2)}\hspace*{1pt} .
\end{equation}
For any value\, $\nu \in (0, 0.5]$\, we have\, $(1 + \nu) (1 - \nu^2) > 1$.
Hence\, $\bar{E} < E$\, which in fact means that now the Gao beam is softer than EB beam.

Another inconvenience caused by the coefficient\, $(1 + \nu) (1 - \nu^2)$\, is connected with buckling load values. For EB beam it is well known that this value can be expressed as
\begin{equation}\label{Ecr}
  P^{E\!B}_{cr}\, =\, \min_{v \in V}\hspace*{1pt} \frac{\int_0^\mathrm{L} EI\hspace*{2pt} (v{''})^2 \hspace*{1pt} \mathrm{d}x}{\int_0^\mathrm{L} \hspace*{2pt} (v{'})^2 \hspace*{1pt} \mathrm{d}x}\hspace*{1pt} ,
\end{equation}
where $V$ denotes a space of kinematically admissible deflections. But convexity bound for Gao beam is in some sense connected with the value
\begin{equation}\label{Gaocr}
  \overline{P}\, =\, \min_{v \in V}\hspace*{1pt} \frac{\int_0^\mathrm{L} EI\hspace*{2pt} (v{''})^2 \hspace*{1pt} \mathrm{d}x}{\int_0^\mathrm{L} (1 + \nu) (1 - \nu^2)\hspace*{2pt} (v{'})^2 \hspace*{1pt} \mathrm{d}x}\hspace*{1pt} ,
\end{equation}
see \cite{machalova17}, \cite{machalova18}, and it is evident that\, $P^{E\!B}_{cr} > \overline{P}$,
while for tougher beam this fact is the opposite of what one would expect.

\medskip
In conclusion, with settings of the integral constant $C$ from (\ref{du/dx}) in a form (\ref{C}), (\ref{lambda}) presented in \cite{gao96}, \cite{santos-gao12}, pure bending problems demonstrate that Gao beam is tougher than EB beam but in pure buckling problems the situation is exactly contrary, which is not consistent. In our opinion, the solution of this inconsistency is to set the integral constant $C$ in the form
\begin{equation}
  C\, =\, -\hspace*{1pt} \frac{P\, (1 - \nu^2)}{2\hspace*{1pt} h b \hspace*{1pt} E\hspace*{1pt} (1 + \nu)}\hspace*{1pt} ,
\end{equation}
which leads to the Gao beam equation
\begin{equation}\label{NewGao}
  E I\hspace*{1pt} w{''''}\hspace*{1pt} -\, E \alpha\hspace*{1pt} (w{'})^2 w{''}\hspace*{1pt} +\, (1 - \nu^2)\hspace*{1pt} P \hspace*{1pt} w{''}\hspace*{1pt} =\, (1 - \nu^2)\hspace*{1pt} q.
\end{equation}
Now in pure buckling problem for equation (\ref{NewGao}) after dividing by $(1 - \nu^2)$ we get
\begin{equation}
  \widetilde{E} I\hspace*{1pt} w{''''}\hspace*{1pt} -\, \widetilde{E} \alpha\hspace*{1pt} (w{'})^2 w{''}\hspace*{1pt} +\, P\hspace*{1pt} w{''}\hspace*{1pt} =\, 0,
\end{equation}
which means that the Gao beam is tougher than EB beam also in this case. As a consequence it is now possible to obtain the following inequality for critical (or limit) values of axial load for Gao and EB beam:\, ${P}^{G}_{cr}\, > {P}^{E\!B}_{cr}$.

%%%%%%%%%%%%%%%%%%%%%%%%%%%%%%%%%%%%%%%%%%%%%%%%%%%%%%%%%%%%%%%%%%


\begin{thebibliography}{99}

\bibitem{gao96}
D.Y. Gao: Nonlinear elastic beam theory with application in contact problems and variational approaches, \textit{Mech. Research Communication},  23 (1), pp. 11--17, 1996.

\bibitem{santos-gao12}
H.A.F.A. Santos, D.Y. Gao: Canonical dual finite element method for solving post-buckling problems of a large deformation elastic beam, \textit{Int. J. Nonlinear Mechanics}, 47 (2), pp. 240--247, 2012.

\bibitem{machalova17}
J. Machalov\'{a}, H. Netuka:
Control variational method approach to bending and contact problems for Gao beam.
\textit{Applications of Mathematics}, Vol. 62, No. 6, pp. 661--677, 2017.

\bibitem{machalova18}
J. Machalov\'{a}, H. Netuka:
Solution of Contact Problems for Gao Beam and Elastic Foundation.
\textit{Mathematics and Mechanics of Solids}, Special Issue on Inequality Problems In Contact Mechanics, Vol. 23, Issue 3, pp. 473--488, 2018.

\end{thebibliography}
\end{document}